\documentclass{article}
\usepackage{graphicx} 
\usepackage{amsmath}
\usepackage{natbib}   

\title{
Detecting Spatial Outliers: the Role of the Local Influence Function
}
\author{Giuseppe Arbia, Vincenzo Nardelli}
\date{}

\begin{document}

\maketitle

\begin{abstract}
In the analysis of large spatial datasets, identifying and treating spatial outliers is essential for accurately interpreting geographical phenomena. While spatial correlation measures, particularly Local Indicators of Spatial Association (LISA), are widely used to detect spatial patterns, the presence of abnormal observations frequently distorts the landscape and conceals critical spatial relationships. These outliers can significantly impact analysis due to the inherent spatial dependencies present in the data. Traditional influence function (IF) methodologies, commonly used in statistical analysis to measure the impact of individual observations, are not directly applicable in the spatial context because the influence of an observation is determined not only by its own value but also by its spatial location, its connections with neighboring regions, and the values of those neighboring observations. In this paper, we introduce a local version of the influence function (LIF) that accounts for these spatial dependencies. Through the analysis of both simulated and real-world datasets, we demonstrate how the LIF provides a more nuanced and accurate detection of spatial outliers compared to traditional LISA measures and local impact assessments, improving our understanding of spatial patterns.
\end{abstract}

Some key words: Robust statistics, Empirical influence function, Local indicator of spatial association, local Moran, spatial correlation.

\section{Introduction}

In the world of big spatial data, it is more and more necessary to develop tools that can help summarizing the many geographical features that may be observed in spatial datasets. In this context, the existing global measures of spatial correlation (such as, e. g., the \cite{moran1950notes}  coefficient), very rarely can characterize in an exhaustive way the many different facets of spatial distributions and, to overcome this limitation, local measures have been introduce. Amongst them, an important category is represented by the class of Local Indicators of Spatial Association (LISA), introduced by Luc Anselin \citep{anselin1995local}. Indeed, when analyzing real data, the LISA maps are easy-to-interpret visualization tools that help identifying local phenomena of clustering (such as those associated with significant HH and LL local Moran indicators) and spatial outliers (such as those associated with HL and HL local Moran indicators). In this paper, we approach the study of spatial outliers from a different point of view, and we develop a new method for spatial outlier detection that is alternative (and complementary) to the classical LISA indicators. In particular, we will approach the study of spatial outliers starting from the classical literature on robust statistics \citep{hampel1974influence, rousseeuw2005robust, Huber1981, Hampel1986, Maronna2006}. In this area an important role is played by concepts like the "breakdown point", the "sensitivity curve" and the "influence function" (IF) that can help quantifying how much a statistical measure is unduly affected by the presence of abnormal observations. However, as it is shown in \cite{nardelli2024robust}, when data are distributed in space, the traditional definition of influence function cannot be employed because, due to the intrinsic dependence characterizing spatial data, the influence of an outlier is not independent of its location and it is strongly affected by the spatial context where it is observed. In this paper, we first introduce a local version of the spatial influence function (proposed in \cite{nardelli2024robust}), we then characterize such a function parametrically to facilitate the interpretation and the associated visualization, and we test its effectiveness in describing geographical features through the analysis of simulated and real data. The proposed local influence function (LIF) differs from LISA indicators in several key aspects. While LISA measures focus primarily on identifying local spatial autocorrelation, the LIF complements this by emphasizing the role of individual outliers within the spatial context, accounting not only for their value but also for their influence on neighboring areas. LISA effectively identifies clusters (high-high and low-low) or spatial outliers (high-low and low-high) based on local similarity or dissimilarity in values, but it does not directly measure the extent to which outliers distort the overall spatial pattern. The LIF enhances this analysis by directly measuring the influence of extreme observations, considering both their magnitude and spatial dependencies. Together, LISA and the LIF provide a more comprehensive approach to identifying outliers and assessing their impact on the surrounding spatial structure.

The rest of the paper develops as follows. Section 2 will be devoted to introducing some important definitions and to develop a local version of the spatial influence functions proposed in \cite{nardelli2024robust}. Section 3 has the aim to show the empirical relevance of the proposed new tool through a series of Monte Carlo experiments, while Section 4 illustrates it examining two real datasets. Finally, Sections 5 draws some tentative conclusions and examines some possible future developments.

\section{The spatial influence function and its local version}

In this section we wish to introduce a set of tools that are necessary to assess the robustness of spatial correlation measures. To start with, suppose that we have at our disposal, say, $n$ observations of a random variable Z such that $\mathrm{Z}=\left(z_{1}, z_{2}, \ldots, z_{n}\right)$ centered around their mean and, without loss of generality, with unitary variance). The observations $n$ are distributed on a (possibly irregoular) lattice. The celebrated Moran spatial correlation coefficient (Moran, 1950) can then be defined as follows:

\begin{equation}
MC=\frac{\sum_{i=1}^{n}\left(z_{i}\right) L\left[z_{i}\right]}{\sum_{i=1}^{n}\left(z_{i}\right)^{2}}=\frac{Z^{T} L(Z)}{Z^{T} Z}
\end{equation}

with $L\left(z_{i}\right)=\sum_{i=1}^{n} w_{i j} z_{j}$ the spatial lag, $w_{i j}\left\{\begin{array}{l}=0 \text { if } i=j \\ >0 \text { if } j \in N(i) \\ =0 \text { otherwise }\end{array}\left(w_{i j} \in W\right).\right.$ and with $N(i)$ representing the set of locations connected with location $i$. Following a consolidated tradition, in the reminder of the paper we will further assume that the W matrix is row-standardized so that $\sum_{j=1}^{n} w_{i j}=1$ for each i.

When assessing the robustness of Moran's coefficient, in principle, we could follow the traditional approach introduced by Hampel (1974) and look at the associated empirical influence function. In general, the finite sample version of Hampel's influence function \cite{hampel1974influence} can be defined as $I_{+}\left(\theta, z_{o}\right)=(n+1)\left(\hat{\theta}_{+}-\hat{\theta}\right)$ with $\hat{\theta}$ as an estimator of a generic parameter $\theta$, based on n observations, and $\hat{\theta}_{+}$an estimator of the same form of $\hat{\theta}$ based on the same $n$ observations, but also on one additional observation, say $z_{o}$. However, when used within a spatial context, such a definition cannot be employed because a spatial sample, in general, is constituted by a collection of, say $n$, areas (or points) which are given once and for all (e. g. the region in one country) so that we cannot imagine an empirically relevant situation where we can introduce extra observations (e. g. an extra region) in the dataset. In this case \cite{nardelli2024robust} proposed to use the analogous measure:

\begin{equation}
I_{c}\left(\theta, z_{1}\right)=\mathrm{n}\left(\widehat{\theta}_{c}-\widehat{\theta}\right) \tag{2}
\end{equation}

where and $\hat{\theta}_{c}$ is an estimator of the same form of $\hat{\theta}$, but where one of the units (say unit 1 without loss of generality) is contaminated with abnormal values.

In a spatial context in general, the quantity $I_{c}\left(\theta, z_{1}\right)$ depends not only on the amount of contamination $z_{1}$, but also on the location where the contamination is observed, on its connection with the neighboring locations and on the values observed in the neighbouring locations. In fact, given the nature of spatial dependence between observations, if the contaminated a generic location $i$ is strongly connected with other locations (so that $i$ can be considered a dominant unit according to the definition of \cite{pesaran2020econometric}), the influence of $z_{1}$ will be stronger because, in this case, its disturbing effects propagate also to the neighboring units corrupting more substantially the parameter estimation.

Let us now refer more specifically to the influence function of the MC coefficient introduced in Equation (1) and let us consider, without losing generality, the further hypothesis that the contaminated unit before contamination assumed a value equal to zero. Under these hypotheses \citep{nardelli2024robust} proved that the Moran coefficient after contamination can be expressed as:
\begin{equation}
\widehat{M C}_{\text {c }}=\frac{\frac{1}{n}\left[\sum_{i=2}^{n} \sum_{j=2}^{n} w_{i j} z_{i} z_{j}+2 z_{1} \sum_{i=2}^{n} w_{i 1} z_{i}-n^{-1} z_{1}^{2}\right]}{\frac{1}{n}\left[n+\frac{n-1}{n} z_{1}{ }^{2}\right]}
\end{equation}

so that, as a consequence, the influence function of $M C$ can be expressed as:

\begin{equation}
I_{c}\left(M C, z_{1}\right)=\frac{2 n z_{1}}{\frac{n-1}{n} z_{1}{ }^{2}+n} \sum_{i=2}^{n} w_{i 1} Z_{i}-\frac{z_{1}^{2}}{\frac{n-1}{n} z_{1}{ }^{2}+n}(M C+1)
\end{equation}

Equation (4) shows that the influence function of the Moran coefficient depends on the contaminated value $z_{1}$, as it is obvious, but also on the average of the values observed in the neighbourhood of the contaminated location (the term $\sum_{i=2}^{n} w_{i 1} z_{i}$ ) and on the level of spatial correlation before the contamination (the term $\mathrm{MC})$.

Figure 1 shows the behaviour of the Moran's influence function as an increasing function of both $z_{1}$ and $\sum_{i=2}^{n} w_{i 1} z_{i}$ at 7 different levels of the original MC coefficient ranging from strong negative ( $M C=-0.7$ ) to strong positive ( $\mathrm{MC}=0.07$ ) spatial correlation. The seven maps show that the level of MC only marginally affects the influence function. Furthermore, Figure 2 display the behaviour of the $M C$ influence as a function of $M C$ and $z_{1}$ setting and $\sum_{i=2}^{n} w_{i 1} z_{i}=0$ just for the sake of visualizing the relationship.

\begin{figure}[h!]
    \centering
    \includegraphics[width=1\textwidth]{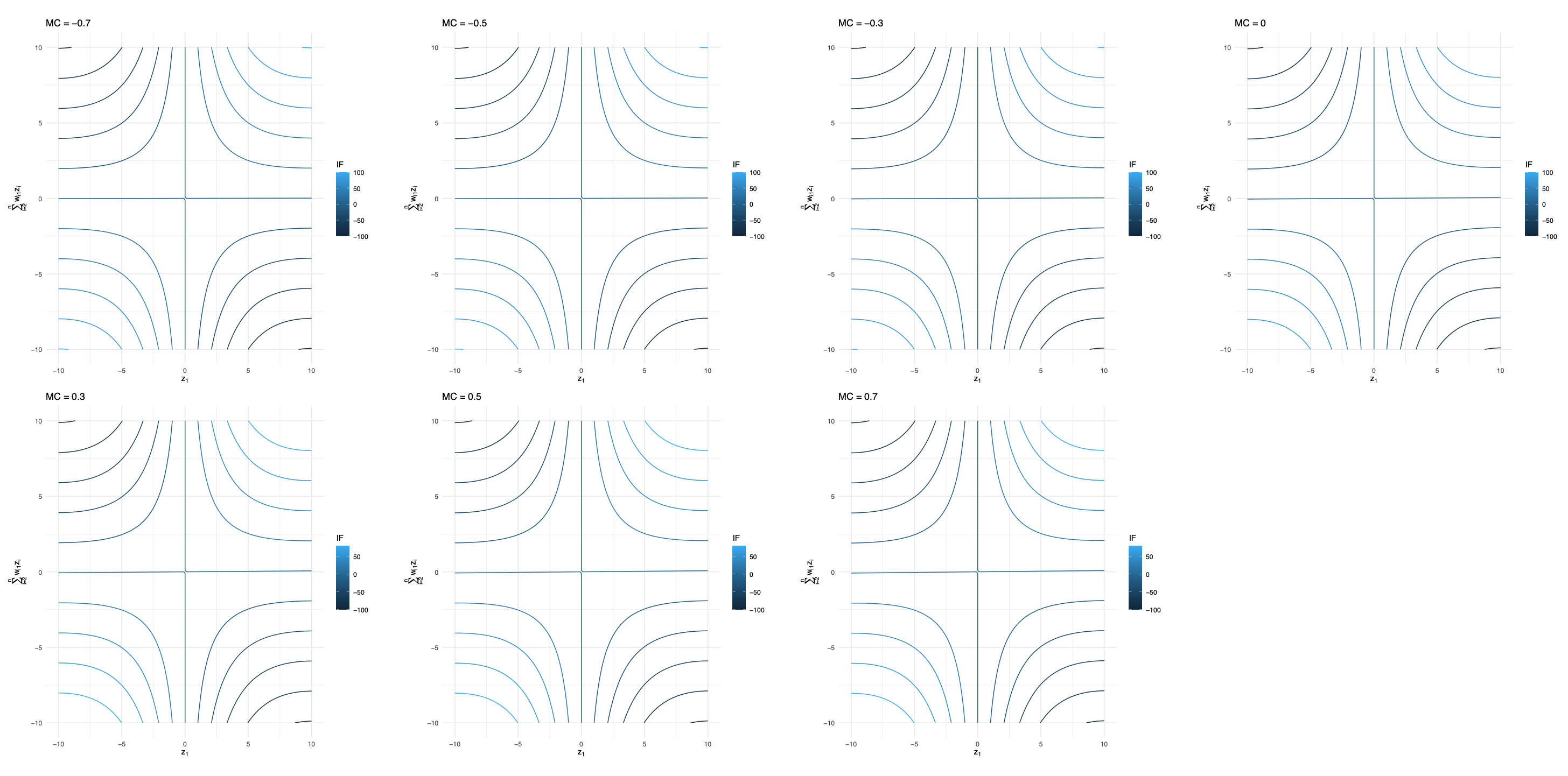}
    \caption{Moran's influence function expressed as function of $z_{1}$ and $\sum_{i=2}^{n} w_{i 1} z_{i}$ for 7 different levels of the original MC coefficient ranging from strong negative (MC=-0.7; top left corner) to strong positive $(M C=0.07bottom right corner)$}; 
    \label{fig:power2} 
\end{figure}

\begin{figure}[h!]
    \centering
    \includegraphics[width=0.8\textwidth]{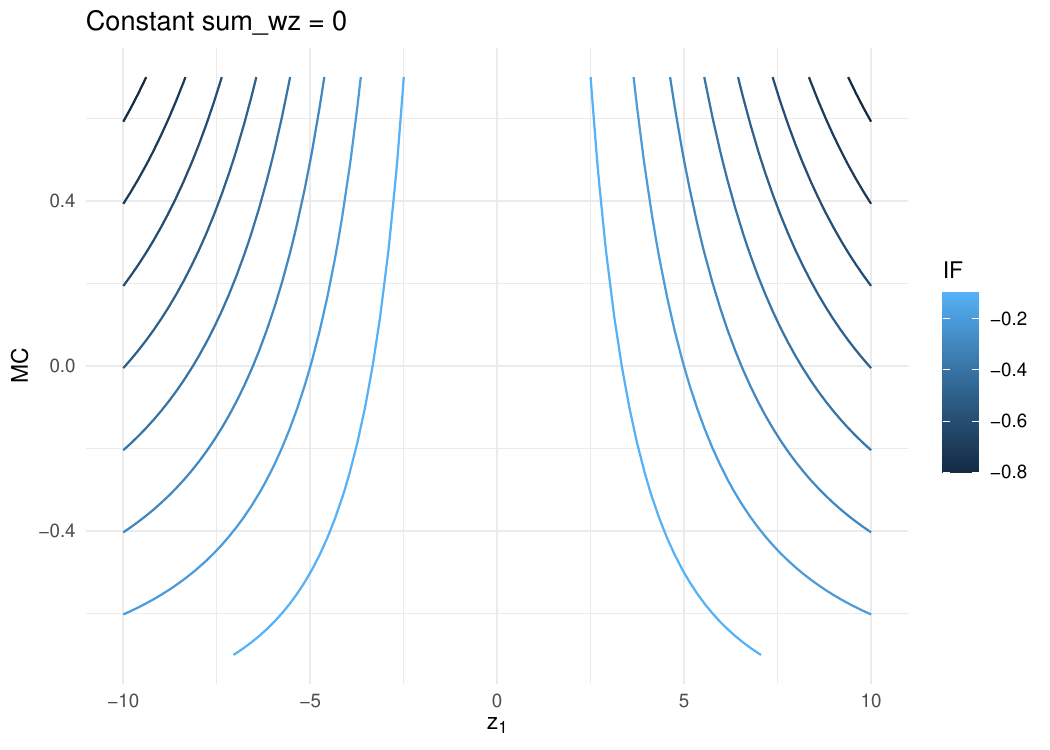}
    \caption{Figure 2: Moran's influence function expressed as function of $z_{1}$ and MC assuming and $\sum_{i=2}^{n} w_{i 1} z_{i}=0$}
    \label{fig:power2} 
\end{figure}

An important consequence of Equation (4) is that the Moran's influence function cannot be represented univocally for a given spatial dataset. On the contrary, in principle, in each of the different location of a map we can observe different levels of the IF related to the spatial location of the contaminated unit. To parametrize this space-varying effect, we propose to characterize each influence function with the integral of its absolute value in the range $[-2 \sigma,+2 \sigma]$:

\begin{align}
\operatorname{LIF}\left(M C, z_{1}\right) &= \int_{-2 \sigma}^{2 \sigma}\left|I_{c}\left(M C, z_{1}\right)\right| d z_{1} \nonumber \\
&= \int_{-2 \sigma}^{2 \sigma}\left|\frac{2 n z_{1}}{\frac{n-1}{n} z_{1}^{2}+n} \sum_{i=2}^{n} w_{i 1} z_{i} - \frac{z_{1}^{2}}{\frac{n-1}{n} z_{1}^{2}+n}(M C+1)\right| d z_{1} \nonumber \\
&= 2\left[n\left(M C+\sum_{i=2}^{n} w_{i 1} z_{i}-1\right) \tan^{-1}\left(\frac{2 \sigma}{n}\right) + 2 \sigma\left(\sum_{i=2}^{n} w_{i 1} z_{i}-1\right)\right] \nonumber \\
&\quad + 2\sqrt{n} \tan^{-1}\left(\frac{2 \sigma}{n}\right)
\end{align}

From now on,we will refer to the expression derived in Equation (4) as to the "Local Influence Function"(LIF) of the MC a measure that can be employed to identifying what are the most influential locations on a map, considering not only the absolute values of the contaminated unit, but also where they are observed and their spatial context. To the aim of illustrating the practical use of the local Influence Function introduced above and its relative merits with respect to other local measures, in the next following sections we will show the results of a series of Monte Carlo experiments (Section 3) and the analysis of different real data sets (Section 4).

\section{Simulation results}

In this section, through the analysis of simulated datasets, we explore the relationship between the spatial distribution of a random variable \( Z \) and the associated Local Influence Function (LIF). In our Monte Carlo experiments, the spatial data were generated on a 10-by-10 regular square lattice grid obeying to a spatial autoregressive process. Specifically, the variable \( Z \) was generated from a normal distribution with mean \( \mu = 0 \), standard deviation \( \sigma = 1 \), and an autocorrelation parameter \( \rho = 0.5 \). The model used for generating the data follows a spatial lag model (SLM), which can be expressed formally as follows:

\[
Z = (I - \rho W)^{-1} \varepsilon
\]

where \( Z \) is the vector of observations, \( \rho \) is the spatial autocorrelation parameter, \( W \) is the (row-standardized) spatial weights matrix representing the neighborhood structure (with inverse \( (I - \rho W)^{-1} \)), and \( \varepsilon \) is a vector of independent and identically distributed errors following a standardized normal distribution:

\[
\varepsilon \sim \mathcal{N}(0, 1)
\]

For the sake of illustrating the simulation process, Figure 3 reports the results of a single realization of the avariable Z on a 10-by-10 regular square lattice grid. In particular, the left pane shows the chorplet map of the variable \( Z \), where darker shades represent low values, and lighter shades indicate high values, highlighting the spatial variation of the phenomenon across the grid. 
In contrast, the choroplet map reported in the right pane of Figure 3 displays the values of the Local Influence Function (LIF). The visual inspection of the graph makes it clear the local contribution of each cell to the overall global spatial autocorrelation (in this case, to Moran’s I coefficient). In particular, cells displaying a high value of the LIF (represented in darker red), indicate regions where local spatial relationships are most influential on the global autocorrelation. These areas correspond to local clusters of similar values (either high or low), where the spatial dependence is strongest.

Thus, the relationship between the spatial distribution of \( Z \) and the LIF is that cells with high LIF values identify regions where the spatial pattern of \( Z \) is most pronounced. These cells highlight local clusters of extreme values, which contribute significantly to the overall spatial autocorrelation as measured by the global Moran’s I.

\begin{figure}[h!]
    \centering
    \includegraphics[width=1\textwidth]{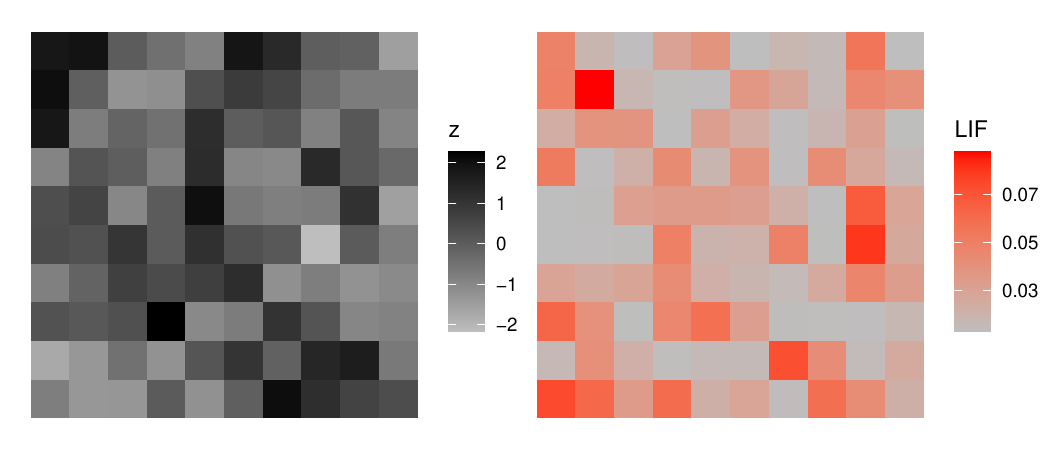}
    \caption{(a) A single realization of the simulated map and (b) Local influence function of the Moran's I}
    \label{fig:power2} 
\end{figure}

The complete results of the simulated maps in 1,000 replications of the simulation are reported in Figure 4, which shows the Local Influence Function (LIF) of Moran's I as it is observed in the two cells where the influence reaches, respectively, its minimum and its maximum values. In the graph reported in Figure 4, the horizontal axis represents the values of the variable \( z_1 \), while the vertical axis represents the Local Influence Function (IF) corresponding to each value of  the contamination\( z_1 \). In particular, the two curves illustrate the behavior of the LIF as it is observed in cell 82 and in cell 96 (cells are numbered progressively from the top-left corner to the bottom right corner).
Cell 82 exhibits maximum local influence on spatial autocorrelation, with the Local Influence Function (LIF) increasing steeply as values deviate from 0. This indicates that extreme values observed in this location significantly impact on Moran's I. In contrast, the LIF observed in Cell 96 demonstrates minimal local influence, with a relatively flat curve indicating a weaker contribution of the values observed in this cell to the global spatial autocorrelation measure.

The complementary nature of LIF and traditional LISA measures is evident in these results. While LISA effectively detects local spatial patterns such as clusters or outliers based on similarity or dissimilarity with neighboring cells, the LIF adds value by directly quantifying the extent to which individual outliers influence the overall spatial correlation. The LIF captures the specific impact of extreme values, taking into account both their magnitude and spatial context. Together, LISA and LIF provide a more comprehensive understanding of spatial patterns, with LISA identifying clusters and LIF offering deeper insights into the broader influence of outliers on the spatial structure.

\begin{figure}[h!]
    \centering
    \includegraphics[width=0.7\textwidth]{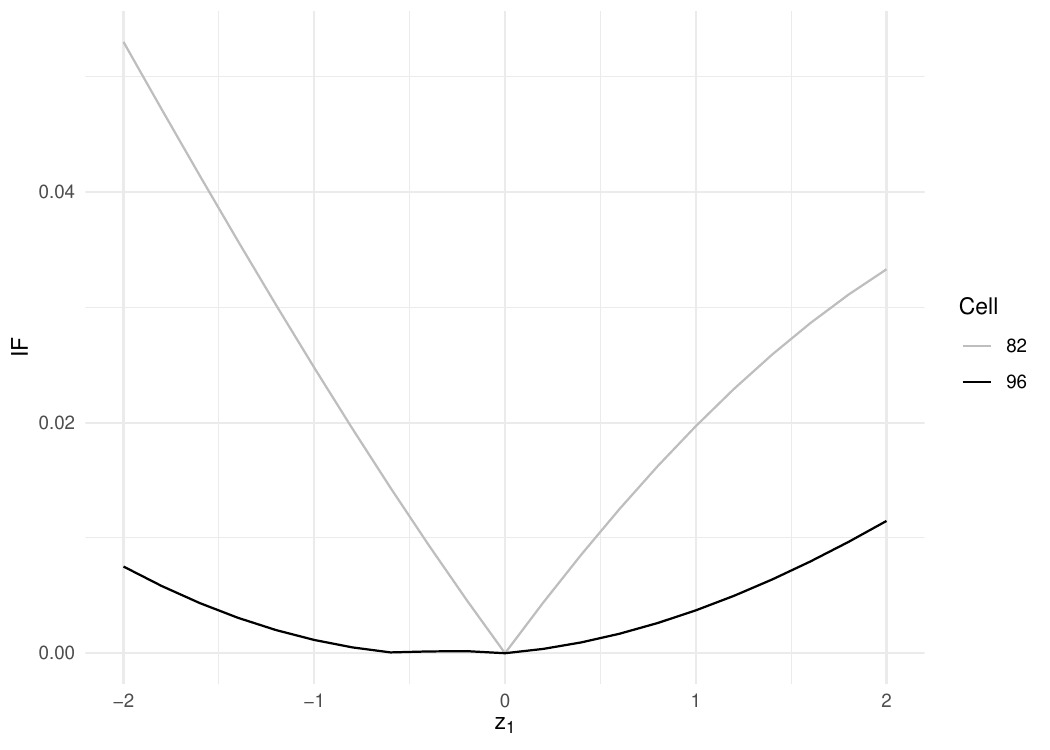}
    \caption{Local influence function of the Moran's I for the two cells where the influence reaches its minimum and maximum.}
    \label{fig:power2} 
\end{figure}

\section{Real Data Analysis}
\subsection{House Prices in Columbus}

Having shown in the previous section the behaviour of the proposed new tool of the Local Influence Function using aseries of  artificially generated datasets, in this section we aim at showing the relevance of the proposed new tool in two real-world situations.
In particular, this Section 4.1 analyses the popular dataset of house prices collected in Columbus, Ohio, while the following Section 4.2 considers some employment data collected by Eurostat, and disaggregated at the NUTS 2
level.

In the first real data analysis, the key variable of interest is represented by the median house value (the variable (\texttt{HOVAL}), observed in each of the 39 neighborhoods of the city of Columbus (Ohio).

The spatial distribution of housing values across Columbus is depicted in Figure 5, which shows significant variability of house prices between the different neighborhoods. Indeed, the housing values range between \$17.90K and \$96.40K, with higher values concentrated predominantly in the northern and central areas, while lower values are clustered in the southern and western areas. This pattern reflects the socio-economic disparities and localized housing market trends within the city. The observed spatial heterogeneity in housing wealth suggests that some neighborhoods are significantly wealthier than others, potentially due to factors such as proximity to economic centers or historical development patterns.

\begin{figure}[h!]
    \centering
    \includegraphics[width=1\textwidth]{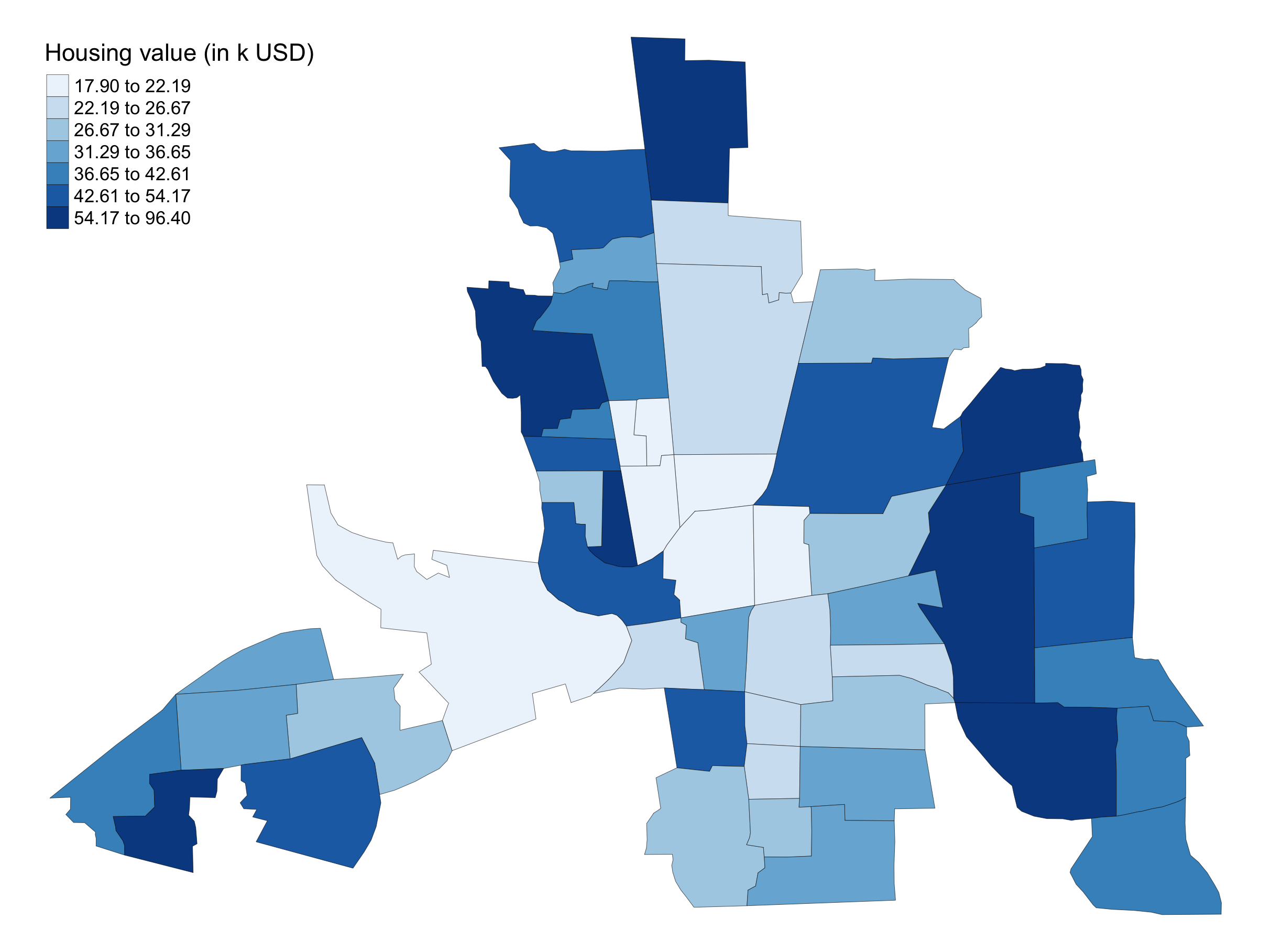}
    \caption{Spatial Distribution of Housing Values in Columbus, Ohio}
    \label{fig:power2} 
\end{figure}

The distribution of \texttt{HOVAL} reveals a notable bimodal pattern, as it is depicted in the density plot  reported in Figure 6(a). The distribution is characterized by two distinct peaks: one centered around \$25K and another less prominent peak at higher values near \$70K. This bimodal distribution suggests the presence of two distinct groups of neighborhoods in Columbus, likely reflecting different socio-economic conditions or housing markets within the city. The lower peak around \$25K indicates a concentration of neighborhoods with relatively modest housing values, while the second peak captures neighborhoods where housing values are significantly higher.

The separation between these two peaks may correspond to distinct geographic or economic divisions within the city. For example, neighborhoods with lower housing values may be clustered in the southern or western parts of the city, as seen in Figure 5, whereas neighborhoods with higher housing values are likely concentrated in the northern and central regions. The bimodal distribution underscores the socio-spatial disparity in housing wealth across Columbus.

Further supporting this observation, the Q-Q plot on Figure 6(b) reveals clear signs of departures from the theoretical normal distribution in both tails. The presence of a larger number of extreme values —both low and high— reinforces the idea that the distribution of house values does not follow a simple normal pattern, but rather a bimodal one. This separation between two groups of neighborhoods, as indicated by the density plot, highlights the polarized nature of the housing market in Columbus.

\begin{figure}[h!]
    \centering
    \includegraphics[width=1\textwidth]{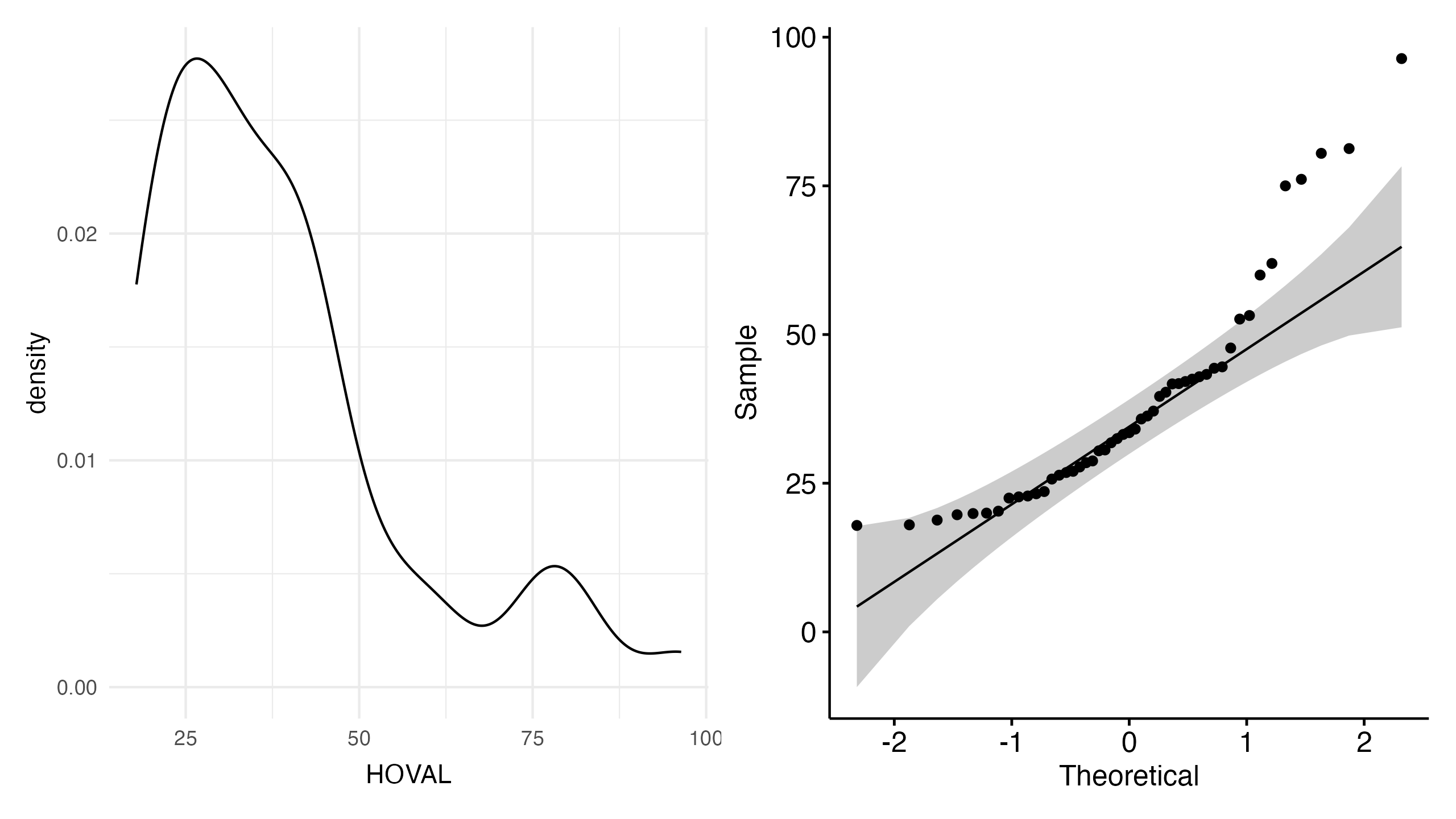}
    \caption{(a) Density plot and (b) QQ Plot of Housing Values (\texttt{HOVAL}) in Columbus, Ohio}
    \label{fig:power2} 
\end{figure}

Figure 7 presents the calculation of the Local Influence Function (LIF) for Moran’s I related to the Columbus dataset and illustrates the local contribution of each neighborhood to the overall spatial autocorrelation of housing values. Indeed, the LIF map identifies key neighborhoods that exert a significant influence on the global spatial pattern of house prices. Neighborhoods with higher LIF values, highlighted in darker shades of red in Figure 7, are those that contribute most to the clustering of housing values. Notably, a northeastern neighborhood stands out with the highest LIF, suggesting that this area plays a dominant role in driving the global spatial autocorrelation of house prices. This could be attributed to its contrast in housing values compared to surrounding areas, enhancing its local influence.

Conversely, neighborhoods with lower LIF values exhibit a more homogeneous distribution of house values, contributing less to the overall spatial autocorrelation. These areas may reflect more gradual transitions in housing prices across space, with less pronounced clusters of high or low values. The LIF analysis thus highlights the spatial heterogeneity of housing values and identifies neighborhoods that are pivotal in shaping the overall spatial structure. This information is crucial for urban planners and policymakers aiming at addressing housing inequalities and at implementing targeted policy measures in areas with distinct spatial dynamics.

\begin{figure}[h!]
    \centering
    \includegraphics[width=1\textwidth]{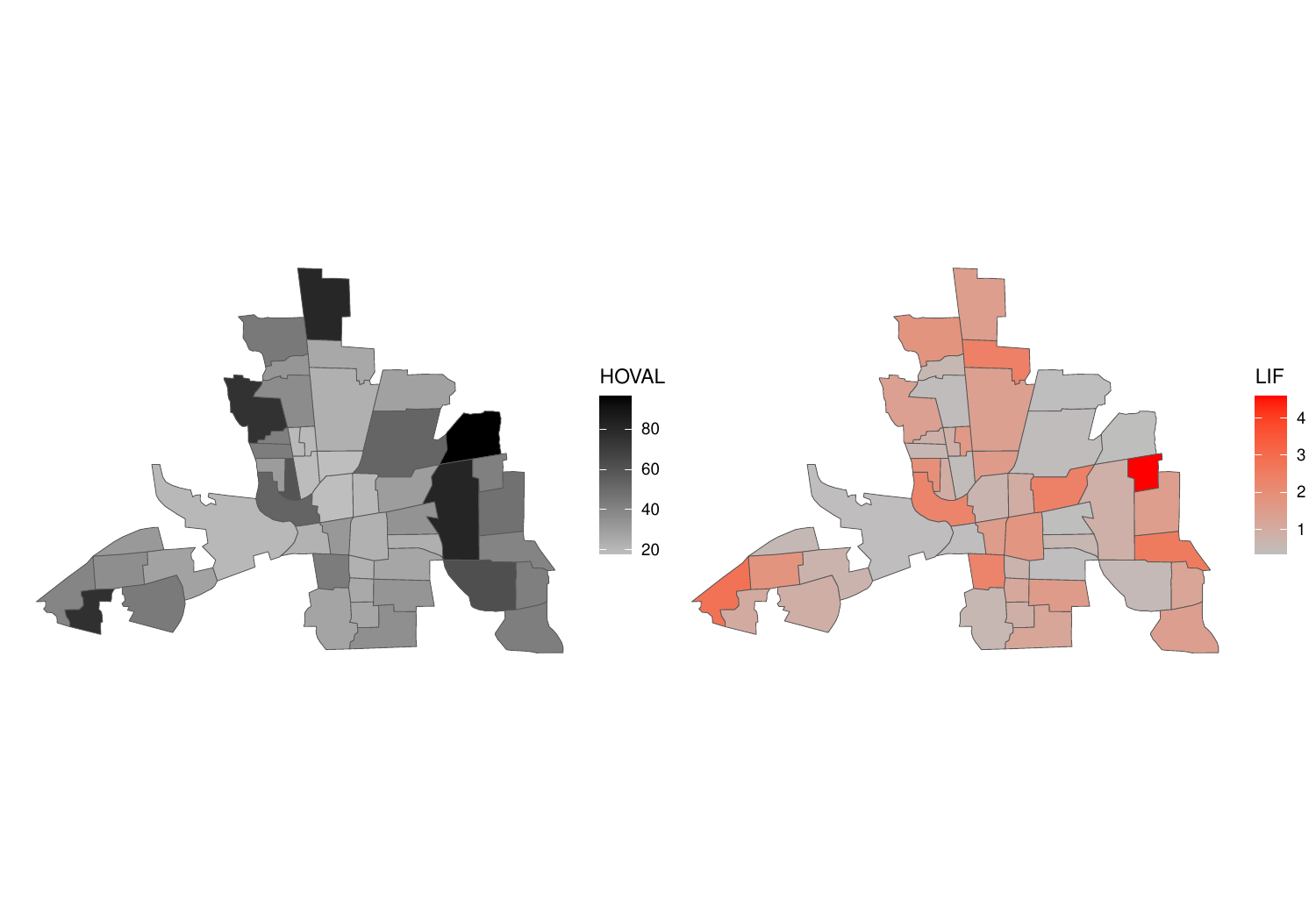}
    \caption{Local Influence Function (LIF) and Spatial Distribution of Housing Values}
    \label{fig:power2} 
\end{figure}

To complement the above results, Figure 8 presents the results of the Local Moran's I (LISA) analysis. The map in Figure 8 (a) highlights, on the left side of the graph, the areas identified as spatial clusters based on significant Local Moran’s I values. Specifically, it shows high-high (HH) clusters in dark red. Additionally, the scatterplot in Figure 8 (b) illustrates the relationship between the lagged values of the variable \texttt{HOVAL} and the corresponding p-values.

Due to the non-normality of the data and the presence of spatial outliers, the Local Moran's coefficients fail to provide meaningful information about spatial dependence across most areas. In fact, the majority of the statistical tests are non-significant at the 0.05 significance level, with only a few neighborhoods exhibiting significant results corresponding to large values of the spatial lag. Notably, these significant p-values are concentrated in a few isolated points, suggesting that the p-values are predominantly driven by extreme values in the spatial lag rather than revealing broader spatial patterns across the entire city.

While LISA effectively identifies clusters and outliers based on local autocorrelation, it may not fully capture the extent to which outliers distort overall spatial relationships. The LIF, in addition, accounts for both the magnitude of outliers and their spatial dependencies, providing a clearer understanding of how specific outliers impact global spatial patterns.

\begin{figure}[h!]
    \centering
    \includegraphics[width=1\textwidth]{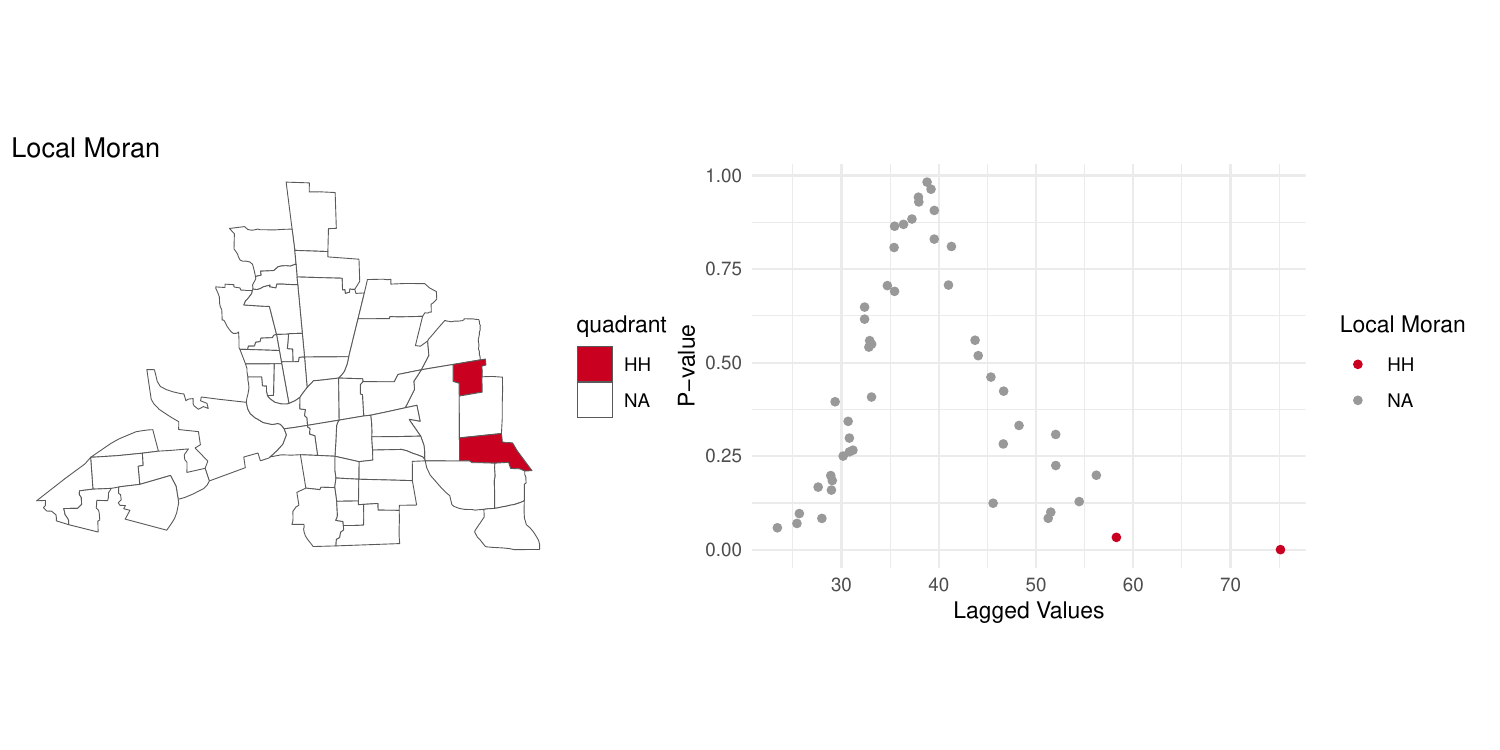}
    \caption{(a) Local Moran’s I (LISA) Analysis and (b) Significance of the spatial clusters}
    \label{fig:power2} 
\end{figure}

When comparing the evidences of Figure 8 to the Local Influence Function (LIF) analysis reported in the previous Figure 7, one can observe that the Local Moran's I analysis identifies only a limited set of areas with significant spatial autocorrelation. In contrast, the LIF provides a more nuanced view of spatial dependencies. Although an area significantly impacts on the others, apparent in both analyses, the LIF can identify additional areas of interest that Local Moran's I does not capture. Thus, the LIF offers a more comprehensive understanding of the spatial dynamics, especially in datasets characterized by non-normal distributions and in the presence of many spatial outliers.

\subsection{Cultural Employment in Europe}
The second case study focuses on a set of cultural employment data derived from the European Union Labour Force Survey (EU-LFS) led by EUROSTAT at the NUTS 2 level. The NUTS 2 regions represent territorial units used for regional statistics across the European Union and correspond to a regional subnational level. It should be clarified that, according to the Eurostat nomnclature, cultural employment includes all individuals engaged in economic activities classified as cultural under the Statistical Classification of Economic Activities in the European Community (NACE Rev. 2) and the International Standard Classification of Occupations (ISCO). This definition includes all people working in cultural activities irrespective of their occupation.

\begin{figure}[h!]
    \centering
    \includegraphics[width=0.7\textwidth]{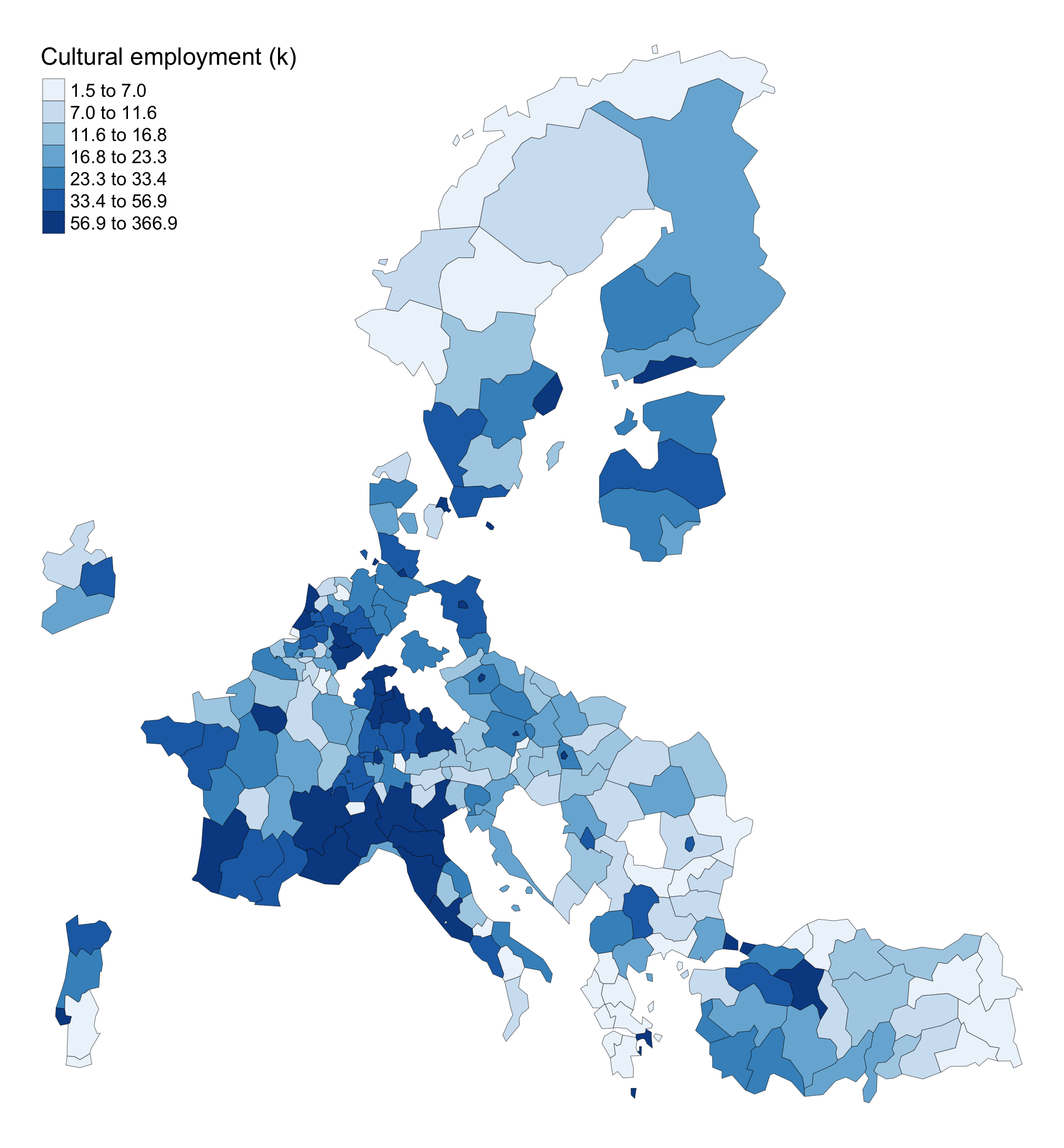}
    \caption{Choroplet map of cultural employment across NUTS 2 regions in Europe.}
    \label{fig:power2} 
\end{figure}

The spatial distribution of cultural employment varies significantly across NUTS 2 regions in Europe, as it is shown in Figure 9. Urban and economically developed regions, particularly in France, Germany and Italy, demonstrate higher concentrations of cultural employment, while northern European regions tend to have lower levels of employment in the cultural sector. This uneven distribution highlights disparities in the availability of jobs in the cultural sector often associated with regional economic and infrastructure development.

The distribution of cultural employment is highly right-skewed, as it is clearly illustrated in Figure 10, with the majority of NUTS 2 regions having relatively low levels of employment, while a small number of regions report exceptionally high values. Such a  skewed distribution is visually represented in the density plot reported in Figure 10 (a), where a pronounced peak is observed for regions with low employment, tapering into a long tail for regions with higher employment levels. Furthermore, the Q-Q plot reported in Figure 10 (b), confirms the such a feature, showing that extreme high values deviate from the expected normal distribution, thus indicating that cultural employment is disproportionately concentrated in a few key regions.

\begin{figure}[h!]
    \centering
    \includegraphics[width=1\textwidth]{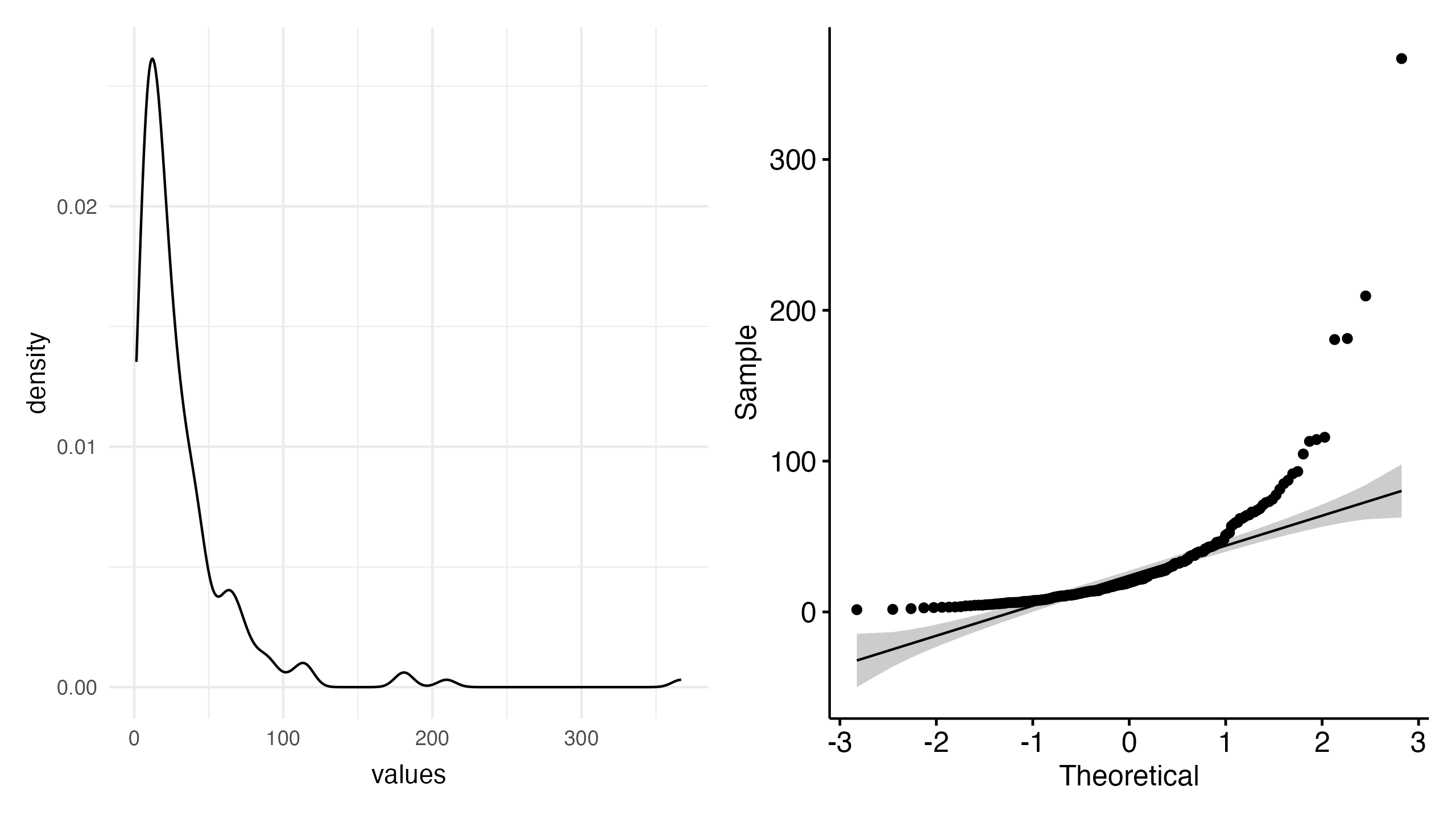}
    \caption{(a) Density plot and (b) Q-Q plot of cultural employment values across NUTS2 regions.}
    \label{fig:power2} 
\end{figure}

The Local Moran’s I analysis reported in Figure 11 (a) reveals only a few significant spatial clusters at the significance level of 0.05, with a dominance of the high-high (HH) patterns. These clusters, where regions with high cultural employment are neighbored by similar regions, are primarily influenced by extreme employment values, as shown by the associated p-values reported in Figure 1 (b). The Local Moran's I generally fails to detect significant spatial autocorrelation in most regions, underscoring the spatial heterogeneity of cultural employment across the NUTS 2 regions.

\begin{figure}[h!]
    \centering
    \includegraphics[width=1\textwidth]{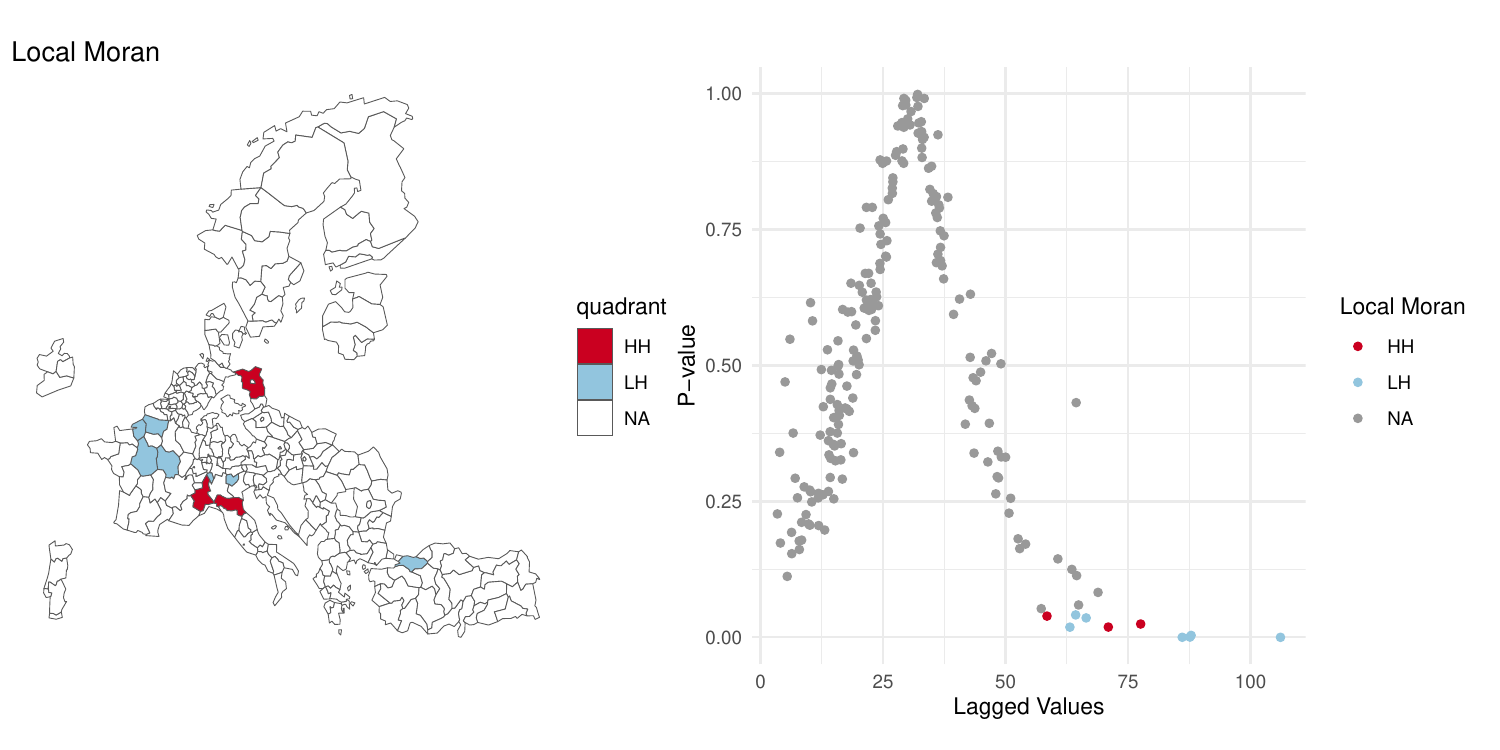}
    \caption{(a) Local Moran’s I (LISA) Analysis and (b) Significance of the spatial clusters for cultural employment across NUTS2 EU regions.}
    \label{fig:power2} 
\end{figure}

The Local Influence Function (LIF) analysis in Figure 12 provides a complementary and more detailed perspective on the spatial structure of cultural employment. While Local Moran’s I identifies only a few significant clusters, the LIF uncovers additional regions that exert moderate influence on overall spatial autocorrelation. This suggests that the LIF is more sensitive to detecting subtler patterns of spatial dependence that Local Moran's I may not capture, allowing for the identification of regions that, while not forming distinct clusters, still play a critical role in the spatial dynamics. By combining LISA with the LIF, analysts gain a more comprehensive understanding of spatial patterns, as LISA highlights clusters and the LIF adds depth by identifying the broader influence of spatial outliers.

\begin{figure}[h!]
    \centering
    \includegraphics[width=1\textwidth]{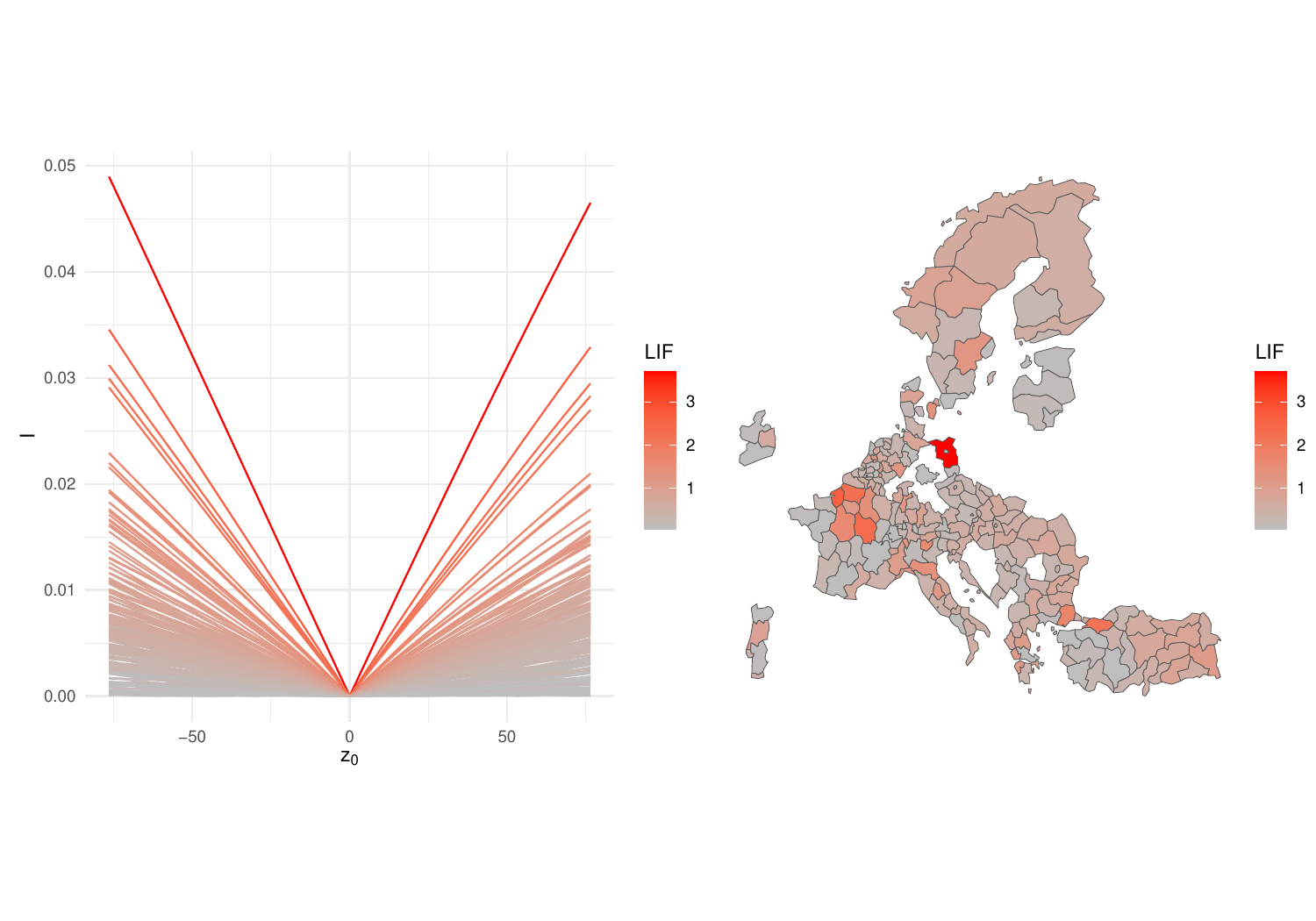}
    \caption{a) Local Moran’s I (LISA) Analysis and (b) Significance of the spatial clusters for the cultural employment across the NUTS2 EU regions.}
    \label{fig:power2} 
\end{figure}

\section{Conclusions}

In this paper, we introduced the Local Influence Function (LIF) as a complementary approach to the commonly used Local Indicators of Spatial Association (LISA) for detecting spatial outliers in large spatial datasets. Our primary focus was on addressing the limitations posed by abnormal observations, which often distort the interpretation of spatial patterns and mask critical spatial relationships. The traditional definition of the Influence Function (IF) popular in the robust statistics literature, was adapted to the spatial context, accounting for the influence of an abnormal observation's value together with its location and its connections to neighboring regions.

Through both simulated and real-world datasets, we demonstrated the utility of the LIF in providing a more nuanced understanding of spatial dependencies. Compared to traditional LISA measures, the LIF proved more effective in identifying spatial outliers and capturing subtler patterns of spatial dependence, especially in cases where non-normal distributions and spatial outliers are present.

Our findings highlight that the LIF is a valuable tool for spatial analysis, offering improved sensitivity in detecting influential regions and spatial outliers. By accounting for both the magnitude of outliers and their spatial dependencies, the LIF provides a more nuanced understanding of how abnormal observations affect spatial autocorrelation and regional dynamics. This has significant implications for fields such as urban planning, regional policy development, and socio-economic analysis, where spatial outliers can critically influence decision-making processes.

Moreover, the LIF complements traditional methods like LISA by uncovering subtler spatial patterns that may be overlooked when only relying on local spatial autocorrelation measures. Together, LISA and LIF provide a comprehensive approach to spatial analysis, with LISA identifying clusters and LIF measuring the broader impact of outliers.

Future research could explore the extension of the concept of the LIF to investigate the impact of abnormal observations on spatial econometric model estimation and testing. This would further expand the utility of LIF in fields that rely on spatial modeling, allowing for a deeper understanding of how outliers influence both spatial relationships and econometric results.

\bibliographystyle{chicago} 
\bibliography{references}

\end{document}